# Effect of Mobility and Traffic Models on the Energy Consumption in MANET Routing Protocols

Said EL KAFHALI, Abdelkrim HAQIQ

*Abstract*— A Mobile Ad hoc Network (MANET) is a group of mobile nodes that can be set up randomly and formed without the need of any existing network infrastructure or centralized administration. In this network the mobile devices are dependent on battery power, it is important to minimize their energy consumption. Also storage capacity and power are severely limited. In situations such as emergency rescue, military actions, and scientific field missions, energy conservation plays an even more important role which is critical to the success of the tasks performed by the network. Therefore, energy conservation should be considered carefully when designing or evaluating ad hoc routing protocols. In this paper we concentrated on the energy consumption issues of existing routing protocols in MANET under various mobility models and whose connections communicate in a particular traffic model (CBR, Exponential, and Pareto). This paper describes a performance comparison of the AODV, DSR and DSDV routing protocols in term of energy consumed due to packet type (routing/MAC) during transmission and reception of control packets. The mobility models used in this work are Random Waypoint, Manhattan Grid and Reference Point Group. Simulations have been carried out using NS-2 and its associated tools for animation and analysis of results.

## I. INTRODUCTION

A Mobile Ad-hoc Network is a group of two or more devices or nodes or terminals with wireless communications and networking competence that communicate with each other without the help of any centralized administrator [9]. Also the wireless nodes that can form a network to exchange information according to their need at that time are an infrastructure less network of mobile devices connected by wireless links. The nodes in this type of networks are generally power constrained because they depend on limited battery resources, whereas wireless communications consume a lot of energy. Without the resource, power, mobile devices will become useless. So, maximizing the lifetime of batteries of each host and entire network is an important issue, especially for MANET, which is supported by batteries only.

Routing packets is one of the main problems in Mobile Ad-hoc Network. In order to facilitate communication within the network, a routing protocol is used to discover routes between nodes. The primary goal of such a MANET routing protocol is correct and efficient route establishment between a pair of nodes so that messages may be delivered in a timely manner. Although establishing efficient routes is an important goal, a more challenging goal is to provide energy consumption routing protocols, since a critical limiting factor for a mobile node is its operation time, restricted by battery capacity. However, the wireless link-only routing path in a MANET makes energy savings difficult to achieve. The corresponding reduction of nodes' lifetime directly affects the network lifetime since mobile nodes themselves collectively form a network infrastructure for routing in a MANET.

In mobile ad hoc network, energy consumption is an important issue as most mobile host operates on limited battery resources. Conservation energy is, therefore, critical in order to prolong the lifetime of the network. There are two main consumers of energy on a MANET node, namely, the central processing unit and the radio (transmitter/receiver). A mobile node not only consumes its battery energy when it is actively sending or receiving packets, but it also consumes battery energy when idle and listening to the wireless medium for any possible communication requests from other nodes. Thus, energy efficient routing protocols minimize either the active communication energy which is required to transmit and receive data packets or the energy consumed during inactive periods. In terms of protocols that belong to the former category, the active communication energy may be reduced by adjusting the radio power of each node just enough to reach the receiving node, and no more. Generally proactive protocols consume more energy due to large routing over heads and reactive protocols suffer from route discovery latencies.

The mobility model plays a very important role in determining the protocol performance in MANET. Thus, it is essential to study and analyze various mobility models and their effect on MANET protocols. In our first work [23], we simulated AODV, DSR and DSDV routing protocols using Manhattan Grid Mobility Model and their performances are analyzed in terms of Packet Delivery Fraction (PDF), Average end-to end Delay and Throughput, in different environments specified by varying network load, mobility rate and number of nodes. In the present analysis, we compare the energy consumption of these protocols under different mobility models using CBR, Pareto and Exponential traffic

Manuscript received on February 2013
**Said EL KAFHALI**, Computer, Networks, Mobility and Modeling laboratory/ Department of Mathematics and Computer/ FST, Hassan 1st University, Settat, Morocco/ E-NGN Research group, Africa and Middle East.
**Abdelkrim HAQIQ**, Computer, Networks, Mobility and Modeling laboratory/ Department of Mathematics and Computer/ FST, Hassan 1st University, Settat, Morocco/ E-NGN Research group, Africa and Middle East.

**Effect of Mobility and Traffic Models on the Energy Consumption in MANET Routing Protocols**

models. The main aim of this paper is to determine the combination of routing protocol, traffic model and mobility model which allows a minimum of energy consumption with various average speeds.

The remainder of this paper is organized as follows: section II reviews the related work. In section III we give a brief description of the studied routing protocols. Section IV presents the mobility models. Section V presents the details of the simulation tools and environments. Simulation results and analysis are described in section VI. Finally, section VII presents our conclusions.

## II. RELATED WORK

The research focus in MANET, in the last few years, has been on developing strategies for reducing the energy consumption of the communication subsystem and increasing the lifetime of the nodes. For examples, authors in [28] have presented the approach called Enhanced Intrusion Detection System (EIDS) for detecting malicious node and minimizing the energy consumption of the node in MANET. This approach leads to less conservation and less communication breakage in ad hoc routing and the experimental results demonstrate that the proposed approach can effectively detect malicious nodes. The authors in [13] have developed the scheme called Energy Based Routing Algorithm (EBRA); they integrated the Dynamic Source Routing (DSR) protocol to ensure the minimum energy consumption rate. The proposed scheme consists of three phases: nodes energy consumption is limited with the high mobility; the effect of malicious behaviour is reduced to avoid the replaying of packets and the unauthenticated node is identified using the digital signature verification. The simulation results show that the proposed scheme achieves less energy consumption rate, more energy efficiency, better throughput, less overhead and delay in the presence of the malicious nodes than the existing schemes. In [19], the authors have evaluated the performance of DSDV, DSR and AODV routing protocols with respect to energy consumption indicating their usage of node's energy considering nodes density and mobility. A new approach for optimizing power consumption in MANETs that consents to maximum life time of mobile hosts while transmitting a packet from the source to destination has been proposed by the authors in [25]; the proposed approach is implemented by introducing a threshold value on each node and transmitting the equal length of packet on the route. The simulation results presented verify the effectiveness of the proposed approach. The authors in [11] have identified the packets responsible for increasing energy consumption with routing protocols using different traffic models. A comparison of the energy consumption of various protocols under CBR traffic was the subject of work in [3]. In [21], the authors have compared the energy consumption of two reactive protocols (AODV and DSR) under Pareto and Exponential traffic. Total energy consumed by each node during transmission and reception process has been evaluated as the function of pause time, speed, number of nodes, and number of sources, sending rate and area shape. The authors in [15] have compared two reactive protocols under ON/OFF source traffic. They have selected packet delivery ratio, normalized routing overhead and average delay as the performance parameters. The authors in [24] have evaluated the energy consumption packets in traffic models (CBR and Exponential) using routing protocols namely AODV and DSDV with parameter variation: number of nodes, pause time, average speed. In [16], the authors have compared the DSDV, AODV and DSR with existing mobility models used in the simulation of MANETs such as Random waypoint, Manhattan Grid, Gauss-Markov, Reference point Group and Heterogeneous Mobility Models.

An analysis of these studies shows that their common goal is to improve the energy consumption of routing protocols. However, the parameters taken into consideration by each of them are different. Some consider only the mobility parameter without addressing the traffic model. Others are interested in the effect of traffic model without taking into account the mobility model, or comparing protocols that belong to the same category (reactive or proactive).

The work presented here compares the energy consumption of protocols AODV, DSR (reactive) and DSDV (proactive) under three models of mobility and three traffic models used at the same time.

## III. ROUTING PROTOCOLS

In recent years, many routing schemes have been proposed. Typically, there are two main categories of routing schemes, proactive schemes and reactive schemes. In the proactive schemes, each node periodically sends control packets to the network in order to maintain a routing table. In the reactive schemes, each node sends control packets for route discovery to find the path to the destination only if they are needed, on demand. A large number of routing protocols have been developed for mobile Ad-hoc networks. In this section, we briefly review the main concepts regarding the three protocols we analyzed, respectively the AODV, DSR and DSDV.

### A. Ad Hoc on-Demand Distance Vector Routing (AODV)

The AODV is a reactive routing protocol for MANETs and other wireless ad-hoc networks provides on-demand route discovery [7]. Reactive routing protocol is meaning that it establishes a route to a destination only on demand. Whenever the nodes need to send data to the destination, if the source node doesn't have routing information in its table, route discovery process begins to find the routes from source to destination. A node requests a route to a destination by broadcasting an RREQ message (Figure 1) to all its neighbours. RREQ message comprises broadcast ID, two sequence numbers, and the addresses of source and destination and hop count. The intermediary nodes which receive the RREQ message could do two steps: If it isn't the destination node then it'll rebroadcast the RREQ packet to its neighbours. Otherwise it'll be the destination node and then it will send a unicast replay message, route replay (RREP) (Figure 2), directly to the source from which it was received the RREQ message. This RREP is unicast along the reverse-routes of the intermediate nodes until it reaches the original requesting node. This process repeats until the RREQ reaches a node that has a valid route to the destination.

At each node, AODV maintains a routing table [26]. Each node has a sequence number. When a node wants to initiate route discovery process, it includes its sequence number and the most fresh sequence number it has for destination. The intermediate node that receive the RREQ packet, replay to the RREQ packet only when the sequence number of its path is larger than or identical to the sequence number comprised in the RREQ packet. A reverse path from the intermediate node to the source forms with storing the node's address from which initial copy of RREQ. Thus, at the end of this request response cycle a bidirectional route is established between the requesting node and the destination. When a node loses connectivity to its next hop, the node invalidates its route by sending an RERR to all nodes that potentially received its RREP [20].

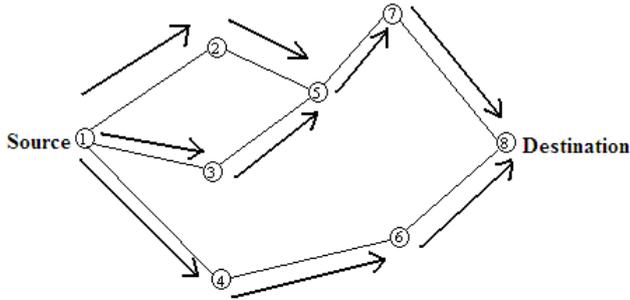

Figure 1: Route Request Message.

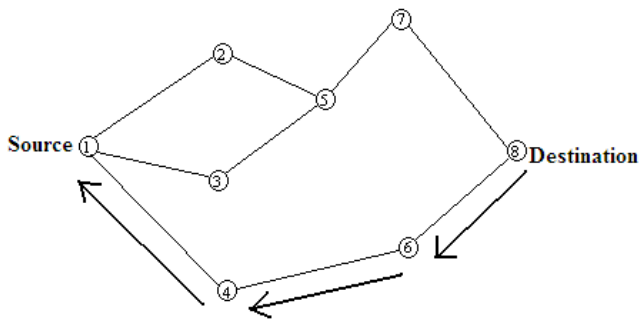

Figure 2: Route Reply Message.

As long as the route remains active, it will continue to be maintained. A route is considered active as long as there are data packets periodically travelling from the source to the destination along that path. Once the source stops sending data packets, the links will time out and eventually be deleted from the intermediate node routing tables. When a source node wants to send data to some destination, first it searches the routing table; if it can find it, it will use it. Otherwise, it must start a route discovery to find a route [2].

*B. Dynamic Source Routing (DSR)*

The DSR is a reactive routing protocol designed specifically for use in multi-hop wireless ad hoc networks of mobile nodes [10]. In this protocol each source determines the route to be used in transmitting its packets to selected destinations. There are two main components, called Route Discovery and Route Maintenance. Route Discovery (Figure 3) is the mechanism by which a node wishing to send a packet to a destination obtains a path to the destination. Route Maintenance (Figure 4) is the mechanism by which a node detects a break in its source route and obtains a corrected route. The sender knows the complete hop by hop route to the destination. These routes are stored in a route cache [5]-[18]. The protocol allows multiple routes to any destination and allows each sender to select and control the routes used in routing its packets, for example for use in load balancing or for increased robustness. The DSR protocol is designed mainly for mobile ad hoc networks of up to about two hundred nodes, and is designed to work well with even very high rates of mobility.

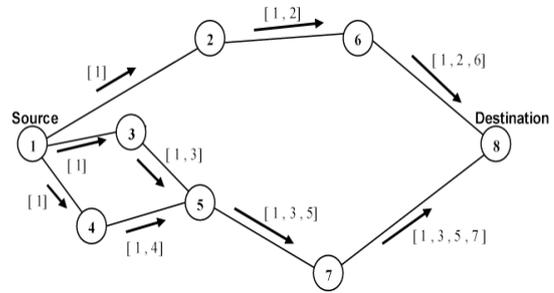

Figure 3: Route Discovery.

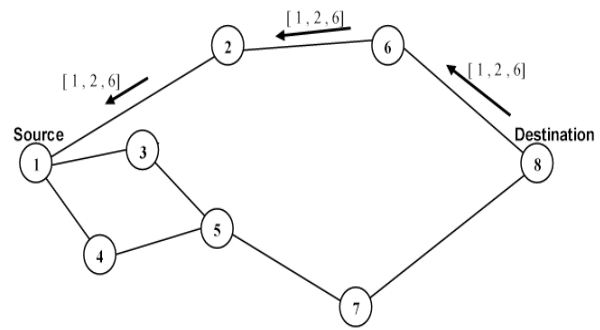

Figure 4: Route Maintenance.

*C. Destination Sequenced Distance Vector (DSDV)*

The DSDV is a proactive routing protocol based on the Bellman-Ford routing algorithm to find the routes with improvements [8]. It was developed by the authors of [22] in 1994. This protocol adds a new attribute, sequence number, to each route table entry at each node. Each node in the mobile network maintains a routing table in which all of the possible destinations within the non-partitioned network and the number of routing hops to each destination are recorded. In this protocol, packets are routed between nodes of an ad hoc network using routing tables stored at each node. Each routing table, at each node, contains a list of the addresses of every other node in the network. Along with each node's address, the table contains the address of the next hop for a packet to take in order to reach the node. This protocol was motivated for the use of data exchange along changing and arbitrary paths of interconnection which may not be close to any base station.

IV. MOBILITY MODELS

The mobility model is designed to describe the movement pattern of mobile users, and how their location, velocity and acceleration change over time. Since mobility patterns may play a significant role in determining the protocol performance, it is desirable for mobility models to emulate the movement pattern of targeted real life applications in a reasonable way. Thus, when evaluating MANET protocols, it

# Effect of Mobility and Traffic Models on the Energy Consumption in MANET Routing Protocols

is necessary to choose the proper underlying mobility model. Mobility models are based on setting out different parameters related to node movement. Basic parameters are the starting location of mobile nodes, their movement direction, velocity range, speed changes over time. The authors in [6] provide a comprehensive survey of mobility models used in simulating Ad-hoc networks. Mobility models are divided into two categories: entity mobility model and group mobility model [27]. Entity mobility model specifies individual node movement. Group mobility model describes group movement as well as individual node movement inside groups. In this work, we consider three mobility models that are designed to capture a wide range of mobility patterns for ad-hoc applications. These models are briefly described in the following subsections.

### A. Random Waypoint Model

Random Waypoint (RW) is a model in which nodes move independently to a randomly chosen destination with a randomly selected velocity [6] (figure 5).

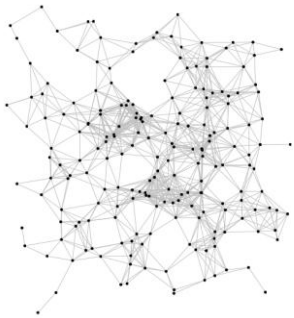

Figure 5: Random Waypoint Mobility Model.

RW includes pause times between changes in direction and/or speed. Pause is used to overcome abrupt stopping and starting in the random walk model. Upon expiry of this pause, the node arbitrary selects a new location to move towards and a new speed which is uniformly randomly selected from the interval [min, max], where min (respectively max) is the minimum (respectively maximum) allowable velocity for every mobile node. After reaching the destination, the node stops for a duration defined by the 'pause time' parameter. After this duration, it again chooses a random destination and repeats the whole process again until the simulation ends. The simplicity of Random Waypoint model may have been one reason for its widespread use in simulations.

### B. Reference Point Group Model

The authors in [14] proposed the group mobility model Reference Point Group Mobility (RPGM) which describes nodes moving in Group. RPGM [6] represents the random motion of a group of mobile nodes and their random individual motion within the group. All group members follow a logical group center that determines the group motion behaviour. The entity mobility models should be specified to handle the movement of the individual mobile nodes within the group. Here, each group has a logical center (group leader) that determines the group's motion behaviour. Initially, each member of the group is uniformly distributed in the neighbourhood of the group leader. Subsequently, at each instant, every node has a speed and direction that is derived by randomly deviating from that of the group leader.

Figure 6 illustrates that each node has a reference point RP(t) within a certain range from the group center which is moved together with the movement of the group center [1].

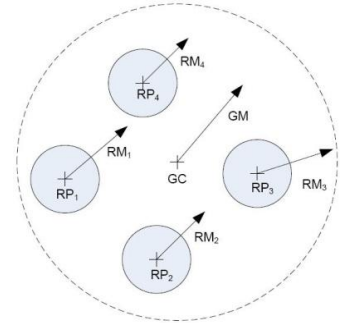

Figure 6: Reference Point Group Mobility Model.

### C. Manhattan Grid Model

Manhattan Grid (MG) has originally been developed to emulate the Manhattan street network, i.e. a city section which is only crossed by vertical and horizontal streets on an urban map (figure 7) [4].

The Manhattan mobility model uses a grid road topology. This mobility model was mainly proposed for the movement in urban area, where the streets are in an organized manner. This mobility model can be described by the following parameters: mean speed, minimum speed (with a defined standard deviation for speed), a probability to change speed at position update, and a probability to turn at cross junctions. The mobile node is allowed to move along the grid of horizontal and vertical streets on the map. At an intersection of a horizontal and a vertical street, the mobile node can turn left, right or go straight. This choice is probabilistic: the probability of moving on the same street is 0.5, the probability of turning left is 0.25 and the probability of turning right is 0.25.

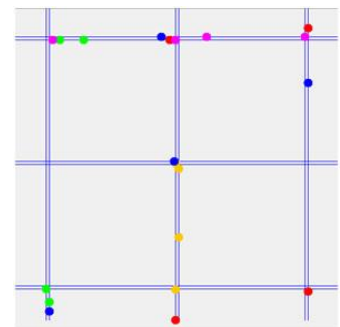

Figure 7: Manhattan Mobility Model.

## V. SIMULATION ENVIRONMENT

### A. Simulation Model

The simulation results presented in this paper were obtained using the NS-2 simulator [12]. This network

245

simulator is a discrete event, object oriented, simulator developed by the VINT project research group at the University of California at Berkeley. We chose a Linux platform i.e. UBUNTU 10.10, as Linux offers a number of programming development tools that can be used with the simulation process. We analyzed the experimental results contained in generated output trace files by using the AWK command. We have generated mobility scenarios for Mobility Models using the BONNMOTION tool [29] and have converted generated scripts to the supported ns2 format so that they can be integrated into TCL scripts.

The simulation parameters are listed in Table 1.

Table 1: Simulation Parameters.

| Parameter | Value |
|---|---|
| Simulator | NS-2 (Version 2.34) |
| Channel type | Channel/Wireless channel |
| Protocols | AODV, DSR and DSDV |
| Simulation duration | 120 second |
| Packet size | 512 kb |
| Traffic rate | 128 bytes |
| Mobility Models | Random Waypoint, Reference Point Group, Manhattan Grid |
| MAC Layer Protocol | 802.11 |
| Traffic Models | CBR, Pareto, Exponential |
| Network size | 50 nodes |
| Topology | 500 m x 500m |

*B. Traffic Models*

Traffic model used in the simulation are CBR, Exponential and Pareto which are generated using cbreng.tcl [17]. Below is a brief description of each traffic:

*B.1 CBR Traffic Model*

CBR generates traffic at a deterministic rate. It is not an ON/OFF traffic.

*B.2 Exponential Traffic Model*

It is an ON/OFF traffic with exponential distribution. It generates traffic during ON period (burst time). Average ON and OFF (idle time) times are 1.5s and 0.5s respectively.

*B.3 Pareto Traffic Model*

It is also an ON/OFF traffic with Pareto distribution. It generates traffic during ON period (burst time). Average ON and OFF (idle time) times are 1.5s and 0.5s respectively with a shape parameter of 2.5.

*C. Energy Consumption Model*

Energy is converted in joules by multiplying power with time. The following equations are used to compute energy required in joules to transmit/receive the packets of given size:

$Energy_{Tx} = (Transmitted\ Power \times Packet\ Size) / 2 \times 10^6$

$Energy_{Rx} = (Receiving\ Power \times Packet\ Size) / 2 \times 10^6$

We have used energy model as given in the following table.

Table 2: Energy Model Parameters.

| Parameter | Value |
|---|---|
| Initial Energy | 150 Joule |
| Idle Power | 1.0 w |
| Receiving Power | 1.1 w |
| Transmission Power | 1.65 w |
| Transition Power | 0.6 w |
| Sleep Power | 0.001 w |
| Transition Time | 0.005 s |

VI. SIMULATION RESULTS AND ANALYSIS

All simulation results show total energy consumed in joule involved in transmitting and receiving the control packets (routing/MAC) with increasing average speed 2m/s, 5m/s, 10 m/s, 15 m/s, 20 m/s and 25m/s.

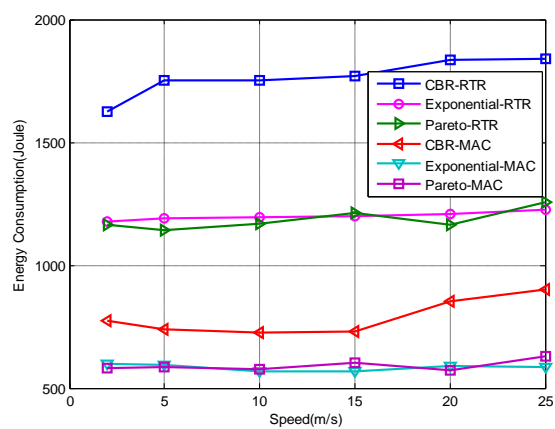

Figure 8: Energy Consumption Vs Speed of AODV protocol with Manhattan Mobility Model.

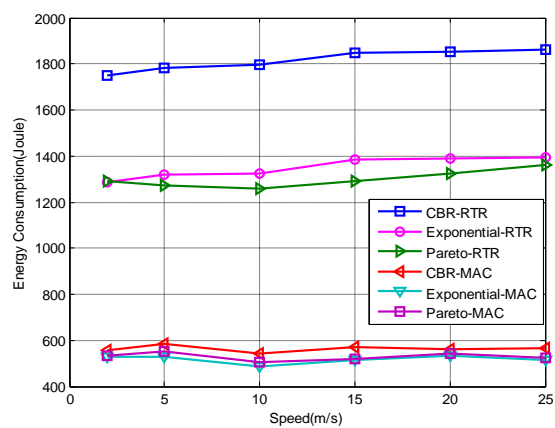

Figure 9: Energy Consumption Vs Speed of DSR protocol with Manhattan Mobility Model.

The performance of the three protocols in different traffic models and under Manhattan mobility model is presented in figures 8, 9 and 10. Similar to DSR, AODV consumes most energy in CBR traffic in comparison of DSDV protocol at routing layer. For other traffic, all protocols consume the same amount of Energy when the average speed increases. At MAC layer, all protocols consume the same amount of energy, except for the fact that the energy consumed by AODV for CBR traffic is higher than Exponential and Pareto traffic.

Figures 11, 12 and 13 show the total energy consumption for the three protocols with RPGM mobility model. In routing

layer, the energy consumption is more with CBR traffic in comparison of Exponential and Pareto traffics for the three protocols. However, with Pareto and Exponential traffics AODV protocol performs low energy consumption in comparison with DSR and DSDV. In MAC layer, all traffic type consume similar amount of energy for all protocols, except for AODV where CBR consumes more energy than other traffics models.

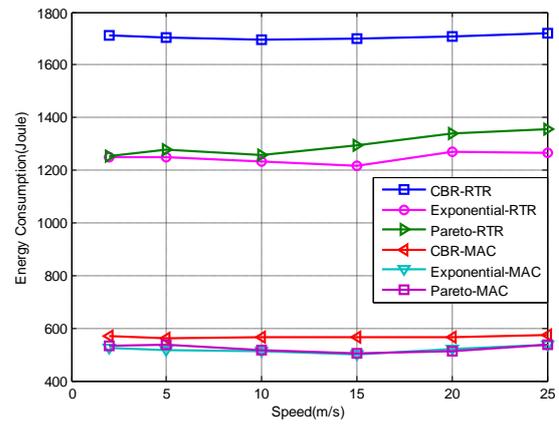

Figure 12: Energy Consumption Vs Speed of DSR protocol with RPGM Mobility Model.

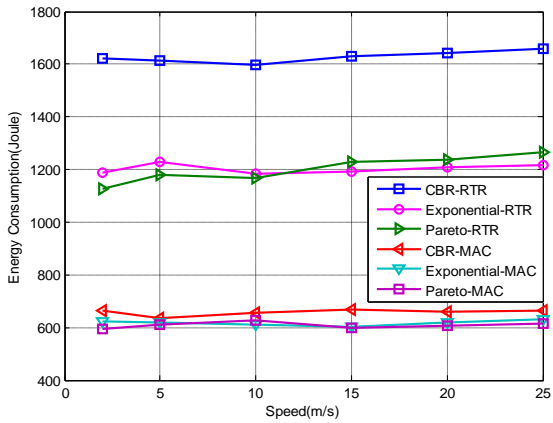

Figure 10: Energy Consumption Vs Speed of DSDV protocol with Manhattan Mobility Model.

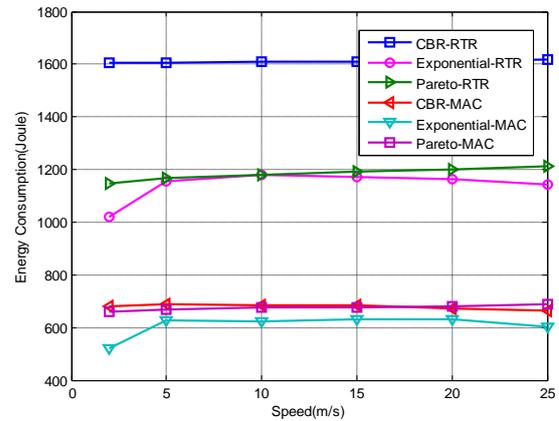

Figure 13: Energy Consumption Vs Speed of DSDV protocol with RPGM Mobility Model.

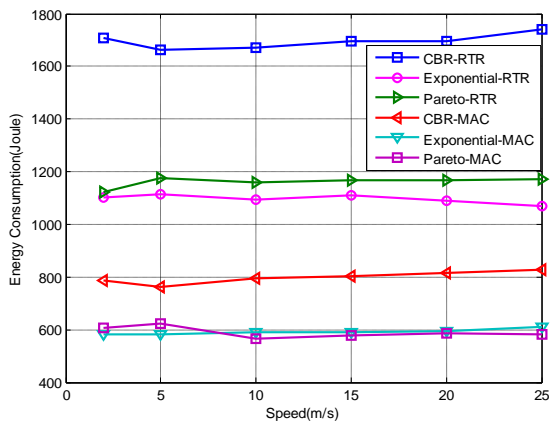

Figure 11: Energy Consumption Vs Speed of AODV protocol with RPGM Mobility Model.

Figures 14, 15 and 16 show the total energy consumption for the three protocols with Random Waypoint mobility model. In the routing layer, the energy consumption is more with CBR traffic in comparison of Exponential and Pareto traffic for the three protocols. However, CBR traffic consumes less energy with DSDV protocol in comparison with the two other protocols. In the MAC layer, with Exponential or Pareto traffic, AODV and DSDV protocols consume more energy than the protocol DSR. Moreover, if we have the CBR traffic it would be more profitable to avoid the AODV protocol.

The figures 10, 13 and 16 confirm that DSDV is not so sensitive to speed and mobility models compared to on-demand protocols, DSR and AODV (figures 8, 9, 11, 12, 14 and 15). The differences among mobility models become subtler here; it is due to the nature of proactive protocols. DSDV works by letting nodes exchange routing tables periodically, therefore the power consumption by this type of routing algorithms tend to stay constant.

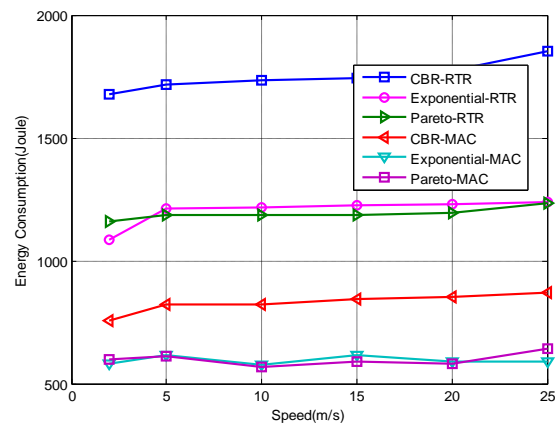

Figure 14: Energy Consumption Vs Speed of AODV protocol with Random Waypoint Mobility Model.

By varying speed level, the topology change is most frequent in Manhattan Grid and Random Waypoint Model than RPGM. In fact, nodes move in groups when RPGM is used as mobility model. This pattern reduces the rate of topology change; that is why it shows uniform and stable energy consumption. Also this model shows lowest amount of energy consumed by every routing protocol. This can be explained by nodes cooperation and reduced topology change. In RPGM model, the amount of power consumed by AODV, DSR and DSDV is almost the same, except when the AODV protocol is used with CBR traffic.

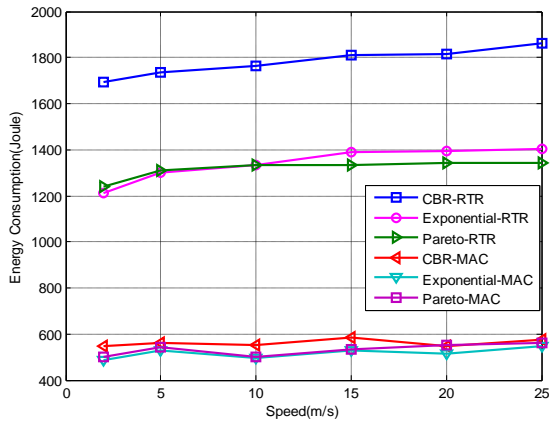

Figure 15: Energy Consumption Vs Speed of DSR protocol with Random Waypoint Mobility Model.

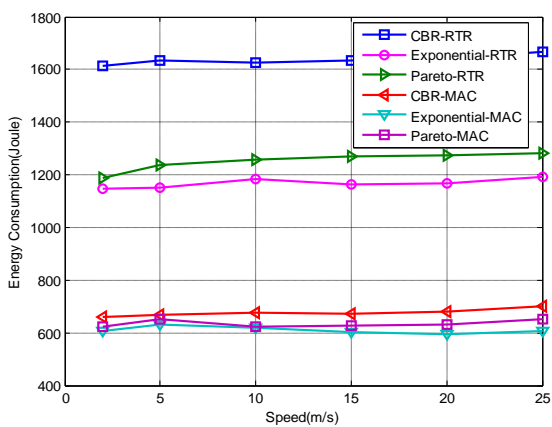

Figure 16: Energy Consumption Vs Speed of DSDV protocol with Random Waypoint Mobility Model.

The simulation results above show that Exponential and Pareto traffic reduce the consumption of energy. This can be explained by their ON/OFF feature.

The obtained results show that AODV and DSR perform better than DSDV at the lowest speed level, but perform worst at high mobility. At very low mobility, the topology changes are less frequent; however, when the speed grows up, routes change and much links are broken forcing AODV to generate much more routing messages (RREQ, RERR…) to look for new routes or to signalize broken links. Therefore, it consumes more energy.

Random Waypoint is considered to be an entirely random scheme and intuitively one would think it may be the most challenging environment for ad hoc routing protocols in terms of energy consumption. The simulation results shown here are consistent with this thought. It can be seen that Random Waypoint model costs the most energy than the others, which infers that a random environment can be more challenging than the other two environments. RPGM with Pareto traffic for AODV protocol is the contrast, the network consumes the smallest amount of energy against Random Waypoint with CBR traffic for AODV or DSR protocols, consumes the largest amount of energy.

MAC layer, the sub-layer of the data link layer, responsible for coordinating and scheduling of transmissions among competing nodes could significantly reduce the power consumption of mobile terminals in MANETs. Indeed, MAC protocols should facilitate the creation of the network infrastructure, these protocols are in charge of fairly and efficiently sharing the wireless channels among a number of mobile terminals and should be energy-aware for extending battery lifetime. All these remarks are confirmed by our simulation results presented here. Our results show that the energy consumption at MAC layer is less for the three protocols with any mobility models and any traffic models in comparison of energy consumption at routing layer. We observed that energy consumed due to MAC control packets significantly affects the total energy consumption for all the three protocols with different mobility and traffic models.

## VII. CONCLUSIONS

In this paper, our study aims to see the impact of mobility and traffic models on the energy consumed by the control packet (routing/MAC) deployed in ad hoc networks. For this, we simulated a network of 50 nodes moving according to a mobility model: Random Waypoint, Manhattan or RPGM, and whose connections communicate in a particular traffic model (CBR, Exponential, and Pareto).

From the above study and obtained simulation results, we observe that with any mobility model and any routing protocols, the network consumes more energy if the traffic used is CBR. By cons, energy consumption varies for other traffic according to the mobility model used and depending on the routing protocol studied. And we show that the energy consumption at MAC layer is less for the three protocols with any mobility models and any traffic models in comparison of energy consumption at routing layer.

We have seen that AODV consume more energy compared to DSR and DSDV with CBR traffic while it consumes less energy compared to DSR and DSDV with Pareto and Exponential traffics. The energy consumption in CBR traffic is more than the Pareto traffic for all mobility models, while energy consumption in Exponential traffic is less than the Pareto traffic.

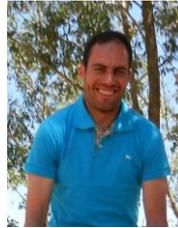

**Said EL KAFHALI** received the B.Sc. degree in Computer Sciences from the University of Sidi Mohamed Ben Abdellah, Faculty of Sciences Dhar El- Mahraz, Fez, Morocco, in 2005, and a M.Sc. degree in Mathematical and Computer engineering from the Hassan 1st University, Faculty of Sciences and Techniques (FSTS), Settat, Morocco, in 2009. He has been working as professor of Computer Sciences in high school since 2006, Settat, Morocco. Currently, he is working toward his Ph.D. at FSTS. His current research interests Communications Network Modelling, Simulation Network Performance, Network Protocols, Mobile Broadband Wireless, and Analysis of Quality of Service in Next Generation Networks.

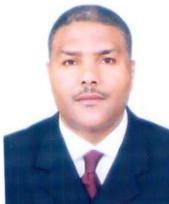

**Dr. Abdelkrim HAQIQ** has a High Study Degree (DES) and a PhD (Doctorat d'Etat), both in Applied Mathematics, option modeling and performance evaluation of computer communication networks, from the University of Mohamed V, Agdal, Faculty of Sciences, Rabat, Morocco. Since September 1995 he has been working as a Professor at the department of Mathematics and Computer at the Faculty of Sciences and Techniques, Settat, Morocco. He is the Director of Computer, Networks, Mobility and Modeling laboratory and the Coordinator of the education of Computer Engineering at the Faculty of Sciences and Techniques, Settat. He is also a General Secretary of e-Next Generation Networks Research Group, Moroccan section.

**Dr. Abdelkrim HAQIQ** is actually Co-Director of a NATO multi-year project entitled "Cyber Security Assurance Using Cloud-Based Security Measurement System" in collaboration with Duke University, USA and Arizona State University, USA. He is also Co-Director of a research project Moroccan Tunisian entitled "Toword T-Learning based on the Smart TV: A Case Study of the Arab Maghreb" in collaboration with University of Sfax, Tunisia.

**Dr. Abdelkrim HAQIQ** is interests lie in the areas of applied stochastic processes, stochastic control, queuing theory, game theory and their applications for modeling/simulation and performance analysis of computer communication networks. He is the author and co-author of more than 50 papers (international journals and conferences/workshops). He was the Chair of the second international conference on Next Generation Networks and Services, held in Marrakech, July, 8- 10, 2010 and the TPC Chair of the fourth international conference on Next Generation Networks and Services, held in Portugal, December, 2 - 4, 2012. He is also an International Steering Committee Chair of the international conference on Engineering Education and Research 2013, iCEER2013, which will be held in Marrakesh, July, 1st –5th, 2013. **Dr. Abdelkrim HAQIQ** is also a TPC member and a reviewer for many international conferences. He is also a Guest Editor of a special issue on Next Generation Networks and Services of the International Journal of Mobile Computing and Multimedia Communications (IJMCMC), July-September 2012, Vol. 4, No. 3.

From January 1998 to December 1998 he had a Post-Doctoral Research appointment at the department of Systems and Computers Engineering at Carleton University in Canada. He also has held visiting positions at the High National School of Telecommunications of Paris, the Universities of Dijon and Versailles St-Quentin-en-Yvelines in France, the University of Ottawa in Canada and the FUCAM in Belgium.